\documentstyle[prl, preprint, aps] {revtex}
\begin{document}

\tightenlines

\title{
Negative Domain Wall Contribution to the Resistivity of\\
Microfabricated Fe Wires}

\author{U. Ruediger, J. Yu, S. Zhang, A. D. Kent}
\address{
Department of Physics, New York University\\
4 Washington Place, New York, New York 10003}

\author{ S. S. P. Parkin}
\address{
IBM Research Division, Almaden Research Center,\\
San Jose, California  95120-6099}

\date{January 8, 1998}

\maketitle

\begin{abstract}
The effect of domain walls on electron transport has been investigated in
microfabricated Fe wires (0.65 to 20 $\mu m$ linewidths) with controlled
stripe domains. Magnetoresistance (MR) measurements as a function of
domain wall density, temperature and the angle of the applied field are
used to determine the low field MR contributions due to conventional
sources in ferromagnetic materials and that due to the erasure of domain
walls. A negative domain wall contribution to the resistivity is found.
This result is discussed in light of a recent theoretical study of the effect
of domain walls on quantum transport.

\end{abstract}

The interplay between the electron transport and magnetic properties of 
ferromagnetic nanowires and thin films has recently been the subject of an 
intense research effort. In mesoscopic ferromagnets an experimental aim has 
been to use magnetoresistance (MR) to study domain wall (DW) dynamics; in 
particular macroscopic quantum tunneling. These experiments have focused 
on the low temperature MR of nanowires of Ni \cite{Hong}, Co, and Fe 
\cite{Otani}. Discontinuous changes in the wire conductance are observed as 
a function of the applied field. These are interpreted as the nucleation and 
movement of DWs which traverse the wire during magnetization reversal. 
In these experiments, nucleation of a DW appears to lead to a decrease in the 
wire's resistivity. Independently a novel theoretical explanation has been 
proposed in which DWs destroy the electron coherence necessary for weak 
localization at low temperature, leading to such a negative DW contribution 
to the resistivity \cite{Tatara}. Another recent experiment suggests large MR 
effects due to DWs can be observed even at room temperature in simple 
ferromagnetic films \cite{Gregg}. A new physical mechanism has been 
proposed to explain these observations which is analogous to that operative 
in giant magnetoresistance (GMR). Within this model the resistivity in the 
presence of DWs is enhanced due to a mixing of minority and majority spin 
channels in a wall in the presence of spin dependent electron scattering 
\cite{Viret,Levy}. This research points to the need for experiments over a 
range of temperatures on ferromagnetic wires with well characterized and 
controllable domain patterns to isolate the important contributions to the MR 
in small samples.  

Here we report on such experiments. Expitaxially oriented micron scale Fe 
wires with controlled domain configurations have been realized to study the 
effect of DWs on magnetotransport properties. In order to isolate the DW 
contribution to the MR the conventional sources of low temperature MR in 
ferromagnets are characterized in detail $-$ both the anisotropic 
magnetoresistance (AMR) and Lorentz MR. As preliminary 
experiments on ferromagnetic nanowires suggest \cite{Hong,Otani}, we find 
that DWs enhance the wire conductance at low temperature. This remarkable 
effect, present in micron scale wires, is difficult to reconcile with the existing 
theories of DW scattering.

The starting point for these experiments are high quality thin (100 nm thick) 
expitaxially grown (110) oriented bcc Fe films. These films display a large 
in-plane uniaxial magnetocrystalline anisotropy, with the easy axis parallel to 
the [001] direction. They are grown on sapphire substrates
as described in Ref.~\cite{Kent}. The films are patterned using projection
optical lithography
and ion milling to produce micron scale wires (0.65 to 20 $\mu$m linewidths of 
$\sim 200\;\mu$m length) and wire arrays (0.65 to 20 $\mu$m linewidths 
of 3 mm length and 10 to 20 $\mu$m spacing)  with the wires 
oriented perpendicular to the magnetic easy axis and parallel to the  [1${\bar 
1}$0] direction. The residual resistivity ratio of 30 and the residual resistance 
$\rho_{o} = 0.2 \; \mu\Omega$cm attest to the high crystalline quality of these 
films.

The competition between magnetocrystalline, exchange and magnetostatic 
interactions results in a pattern of regularly spaced stripe domains 
perpendicular to the wire axis. Varying the wire linewidth changes the ratio 
of these energies and hence the domain size. Fig.~\ref{fig1} shows magnetic 
force microscopy (MFM) measurements of a 2 $\mu$m wire in zero field 
performed at room temperature with a vertically magnetized tip. These 
images highlight the DWs and magnetic poles at the wire edges. For instance, 
clearly visible in Fig.~\ref{fig1}b are light and dark contrast along the DWs 
indicative of Bloch-like walls with sections of different chirality. The 
magnetic domain configurations are strongly affected by the magnetic history 
of the samples. Before imaging the wires were 
magnetized to saturation with a magnetic field transverse (Fig.~\ref{fig1}a) or 
longitudinal (Fig.~\ref{fig1}b) to the wire axis. In the transverse case the 
mean stripe domain length is 1.6 $\mu$m and much larger than in the 
longitudinal case, where it is 0.4 $\mu$m. The observed domain 
structure at $H=0$ is stable over observation times of at least several hours 
showing that the DWs are strongly pinned at room temperature. 

In Fig.~\ref{fig2} the average domain wall separation is plotted as a function 
of wire linewidth and magnetic history. The DW density varies by an order of 
magnitude for the linewidths investigated. Differences between domain 
configurations after transverse and longitudinal saturation are observed for 
wires with linewidths between 1 and 10 $\mu$m. Dotted lines in 
Fig.~\ref{fig1}a illustrate the approximate domain structure. Since current is 
directed along the wire, there are domains with magnetization {\bf M} 
oriented both parallel and perpendicular to the current density {\bf J}. In 
order to estimate the MR contributions due to resistivity anisotropy the 
volume fraction of closure domains (with {\bf M} $\parallel$ {\bf J}) has 
been estimated. Fig.~\ref{fig2} also shows this fraction (labeled $\gamma$) 
determined from MFM images after magnetic saturation in either the 
transverse or longitudinal direction.

MR measurements were performed in a variable temperature high field 
cryostat with in-situ (low temperature) sample rotation capabilities.  The 
applied field was in the plane of the film and oriented either longitudinal 
($\parallel$) or transverse ($\perp$) to the wire axis. A 4 probe ac ($\sim$10 
Hz) resistance bridge with low bias currents (10 to 40 $\mu$A) was employed 
and the magnetic history of the sample was carefully controlled. 
Fig.~\ref{fig3} shows representative MR results on a 2 $\mu$m linewidth 
wire at both  a) high (270 K) and b) low temperature (1.5 K). There is structure 
to the MR in applied fields less than the saturation field ($H_{s\parallel}
= 0.035$ T and $H_{s\perp}= 0.085$T), at which point the 
MR slope changes, and the resistivity then increases monotonically with 
field. At 
270 K the resistivity above the saturation field is larger in the longitudinal 
than in the transverse field orientation, while at 1.5 K this resistivity 
anisotropy is reversed, $\rho_{\perp}(H_s) > \rho_{\parallel}(H_s)$.

Evidently there are competing sources of resistivity anisotropy in these films 
which leads to this reversal of the resistivity anisotropy with temperature.  
Two predominant and well understood sources of low field low temperature 
MR must be considered to interpret this transport data. The first has its 
origins in spin-orbit coupling and is known as AMR $-$ the resistivity 
extrapolated back to zero internal field (B=0) depends on the relative 
orientation of ${\bf M}$ and ${\bf J}$ \cite{Campbell}. The second effect is due to 
the ordinary (Lorentz) magnetoresistance and is also in general anisotropic 
(i.e. dependent on relative orientation of {\bf J}  and {\bf B}) \cite{Schwerer}. 
As Fe has a large magnetization and hence a large internal magnetic field 
($4\pi M=2.2$ T) both factors are of importance. The resistivity of domains parallel 
and perpendicular to the current direction can be written as:
\begin{equation}
\rho_{\perp}(B,T) = \rho_{\perp}(0,T)[1+F_{\perp}(B/\rho_{\perp}(0,T))]
\label{Gperp}
\end{equation}
\begin{equation}
\rho_{\parallel}(B,T) = 
\rho_{\parallel}(0,T)[1+F_{\parallel}(B/\rho_{\parallel}(0,T))]
\label{Gpar}
\end{equation}
Here $B$ is the internal field in the ferromagnet; $B=4\pi M+H-H_d$, with 
$H$ the applied field and $H_d$ the demagnetization field.  The AMR is 
equal to $(\rho_{\parallel}(0,T)-\rho_{\perp}(0,T))/\rho(0,T)$, where $\rho 
(0,T)$ is the average resistivity. The function $F$ is known as the Kohler 
function and parametrizes the ordinary magnetoresistance for longitudinal 
and transverse field geometries in terms of  $B/\rho \sim \omega_c\tau $, 
the cyclotron frequency times the relaxation time\cite{Ziman}. These scaling
functions have 
been determined experimentally by performing MR measurements to large 
fields (6 T) as a function of temperature, as described in Ref. \cite{Schwerer}. 
The scaling relationships (Eqs.~\ref{Gperp} and \ref{Gpar}) are shown in 
Fig.~\ref{fig4}. The inset displays both $\rho_{\perp}(0,T)$ and 
$\rho_{\parallel}(0,T)$ which result from this scaling analysis and which 
overlap on the scale shown. We find $\rho(0,T) \sim aT^2$ with $a=3 \times 10^{-4} 
\mu \Omega cm/K^2$, as typically observed in 3d elemental ferromagnets 
\cite{Campbell}. The AMR is $\sim 4 \times 10^{-3}$ above $80$ K and decreases below 
this temperature. 
The reversal of the resistivity anisotropy at low temperatures ($\rho_{\perp}(H_s) > 
\rho_{\parallel}(H_s)$, Fig.~\ref{fig3}b) is thus
mainly a consequence of the increasing importance of the Lorentz MR 
({\it i.e.} $F_{\perp}^\prime > F_{\parallel}^\prime $). 
At high temperature $\rho(0,T)$ is large 
and $F^\prime(x)_{x \rightarrow 0} \rightarrow 0$, so that the resistivity 
anisotropy is associated with the AMR as seen in Fig.~\ref{fig3}a.

As in all ferromagnetic materials the resistivity anisotropy is a source of low 
field MR. An applied field changes the domain configurations and domains 
with magnetization parallel and perpendicular to the current direction have 
different resistivities. Hence, this low field MR simply reflects the domain 
geometries during magnetization.

There are thus two ways to estimate the DW contribution to the 
resistivity. The first is to perform MR measurements at the temperature at 
which this resistivity anisotropy at $H=0$ vanishes. Since the AMR 
and Lorentz MR contributions to the resistivity anisotropy are of opposite 
sign there will be a temperature at which 
$\rho_{\parallel}(H=0,T_{comp})=\rho_{\perp}(H=0,T_{comp})$, which we 
denote the compensation temperature, $T_{comp}$.  This occurs at 65.5 K and 
MR results are shown in Fig.~\ref{fig5} for a 2 $\mu$m wire. At this 
temperature the low field MR due to the resistivity anisotropy approaches 
zero. However, the measured resistivity at $H=0$ is lower in longitudinal 
than in the transverse field orientation. This correlates with DW density, 
which is larger after longitudinal magnetic saturation (Fig.~\ref{fig1}b). The 
magnitude of the effect also decreases systematically with increasing wire 
linewidth, (Fig.~\ref{fig5}, left-hand inset) and, hence, decreasing DW density 
(Fig.~\ref{fig2}). The observed resistivity at $H=0$ is apparently suppressed in 
the presence of DWs with a magnitude which depends on the density of DWs.

A more definitive correlation between domain configurations, measured at 
room temperature using an MFM, and low temperature MR measurements 
has been established. To do this we warm the sample to room temperature, 
cycle the magnetic field to establish a known $H=0$ magnetic state, and cool. 
The resistivity at $H=0$ and the MR at low temperatures are unchanged for 
these samples in both longitudinal and transverse measurement geometries. 
This is strong evidence that the domain structure is not affected by 
temperature in this range and consistent with temperature dependent 
magnetic hysteresis-loop measurements on wire arrays which show no 
change of the remanent magnetization with temperature.

The temperature dependence of the DW contribution to the resistivity is 
estimated as follows. The effective resistivity in the $H=0$ magnetic state due 
to resistivity anisotropy can be written as \cite{KNote}:
\begin{equation}
\rho_{eff}(H=0,T) = 
\gamma \rho_{\parallel}(B_i,T) + (1-\gamma)\rho_{\perp}(B_i,T)
\end{equation}
where $\gamma$ is the volume fraction of domains oriented longitudinally 
(see Fig.~\ref{fig2}) and $B_i$ is the field internal to these domains ($=4\pi
M-H_d$). We determine $\rho_{\perp}(B_i,T)$ and $\rho_{\parallel}(B_i,T)$  
by extrapolation of the MR data above saturation (again, as indicated by the 
dashed and solid lines in Fig.~\ref{fig5}). The effective resistivity at $H=0$ is 
estimated with the MFM measurements of $\gamma$. Deviations from this 
$\rho_d=\rho(H=0)-\rho_{eff}(H=0)$, i.e., the measured $H=0$ resistivity 
minus this effective resistivity, are negative and depend systematically on 
domain wall density, increasing in magnitude with increasing domain wall 
density. They approach $1.3 \%$ of the wire resistivity at $1.5$ K for a
2 $\mu$m linewidth wire. We also 
find that $|\rho_d|$   decreases with increasing temperature approaching 
zero at $\sim 80$ K (Fig.~\ref{fig5} right-hand inset). This enhancement of 
the conductivity vanishes at $\sim 80 K$ for all the wire linewidths 
investigated.

There are few models of DWs scattering which predict enhancements in the 
conductivity in the presence walls. One is that of Tatara and Fukuyama based 
in weak localization phenomena \cite{Tatara}. They find that DWs contribute 
to the decoherence of conduction electrons which destroys weak localization.
They introduce a wall decoherence time to parametrize this effect 
$\tau_w=\tau/(n_w/(6\lambda k_f^2)(\epsilon_f/\Delta)^2)$. 
Here $\tau$ is the momentum relaxation time, $n_w$ the domain wall 
density, $k_f$ the Fermi vector, $\lambda$ the domain wall thickness, and 
$\epsilon_f/\Delta$ the ratio of the Fermi energy to the exchange splitting of 
the band. With commonly used parameters for s electrons in Fe, 
$\epsilon_f/\Delta \sim 500$, $k_f \sim 1.7 $ \AA$^{-1}$, $\lambda \sim 300$ \AA
, and with $n_w = 2.5 \; \mu$m$^{-1}$ we estimate $\tau_w \sim 
60\tau$. Essential to observing such an effect is the absence of other 
decoherence mechanisms, such as inelastic scattering. Equating 
$\tau_w=\tau_{in}$ gives an upper temperature limit for the presence of 
weak localization phenomena. From the residual resistance $\tau=2.8
\times 10^{-14} s$ and with $\rho_{in}=\alpha T^2$  $(\alpha =3 \times 10^{-4} 
\mu\Omega cm/K^2)$  we find $T_{max}=7$ K.  From this point of view the 
suppression of weak localization due to DWs cannot explain our observation 
of enhanced conductivity up to $\sim 80$ K.

In summary, a new lithographic approach has been used to realize 
ferromagnetic wires with controlled magnetic interactions and hence domain 
configurations. This has enabled a detailed investigation of the low field MR 
in micron scale ferromagnetic wires and, in particular, a study of the effect of 
DWs on the resistivity.  After considering the effects of conventional sources 
of low field MR (AMR and the Lorentz MR), a negative DW contribution to 
the resistivity is identified. While a negative contribution is consistent with a 
recent theory based on weak localization, results above $\sim 10$ K are 
difficult to reconcile with this theory.  Further research of this type, on well 
characterized samples, is clearly warranted to elucidate the interplay between 
the transport and magnetic properties of mesoscopic ferromagnets.

The authors thank Peter M. Levy for helpful discussions of the work
and comments on this manuscript.
This research was supported by DARPA-ONR, Grant \# N00014-96-1-1207.  
We thank C. Noyan for x-ray characterization and M. Ofitserov for technical 
assistance.  Microstructures were prepared at the CNF, project \#588-96.


\begin{figure}
\caption{ MFM images in zero applied field of a 2 $\mu$m linewidth Fe 
wire. Before performing the MFM images the wire was magnetized a) 
transverse and b) longitudinal  to the wire axis.}
\label{fig1}
\end{figure}

\begin{figure}
\caption{The right hand axis displays the domain width versus Fe linewidth
in zero field after transverse (open squares) and longitudinal (open circles)
magnetic saturation.The left hand axis shows the volume fraction of closure domains 
$\gamma$ as function of the linewidth, again, after transverse (
solid squares) and longitudinal (solid circles) magnetization.  }
\label{fig2}
\end{figure}

\begin{figure}
\caption{a) MR data at $270$ K of a 2 $\mu$m wire in the transverse and a 
longitudinal field geometries ($\rho_\perp(H=0,270 K)= 14.7 
\mu \Omega cm$).  b) MR at $1.5$ K again in the longitudinal and transverse 
field geometries ($\rho_\parallel (H=0,1.5 K) = 0.74 \mu \Omega cm$).}
\label{fig3}
\end{figure}

\begin{figure}
\caption{Scaling plot of the transverse and longitudinal MR above magnetic 
saturation for a 2~$\mu$m Fe wire in the form $\rho(B)/\rho(B=0)$ versus 
$B/\rho(B=0)$ at temperatures of (open squares) $1.5 K$, (solid triangles down) 
$40 K$, (open circles) $60 K$, (solid circles) $80 K$, (solid triangles down) $100 K$, 
(solid diamonds) $125 K$, and (open diamonds) $150 K$.  The inset shows the 
scaling parameters $\rho_\parallel(B=0)$ and $\rho_\perp(B=0)$ as a 
function of temperature on a log-log plot.}
\label{fig4}
\end{figure}

\begin{figure}
\caption{MR of a 2 $\mu$m Fe wire at
$65.5$ K. The extrapolation of the high field MR data in transverse (dotted 
line) and longitudinal (solid line) geometry shows that $\rho_\perp (H=0) 
=\rho_\parallel(H=0)$. The resistivity with walls present, $\rho(H=0)$, is 
smaller than this extrapolation and indicates that DWs lower the wire 
resistivity. The left-hand inset shows this negative DW contribution as a 
function of linewidth at this compensation temperature in the longitudinal 
geometry.  The right-hand inset shows the DW contribution as a function of 
temperature deduced using the model described in the text.}
\label{fig5}
\end{figure}


\begin{references}
\bibitem{KNote}
Since the resistivity anisotropy is small the current density 
in each domain is to a good approximation independent of the precise 
domain configurations. We also assume that the domain size is much greater 
than the mean free path.
\end{references}
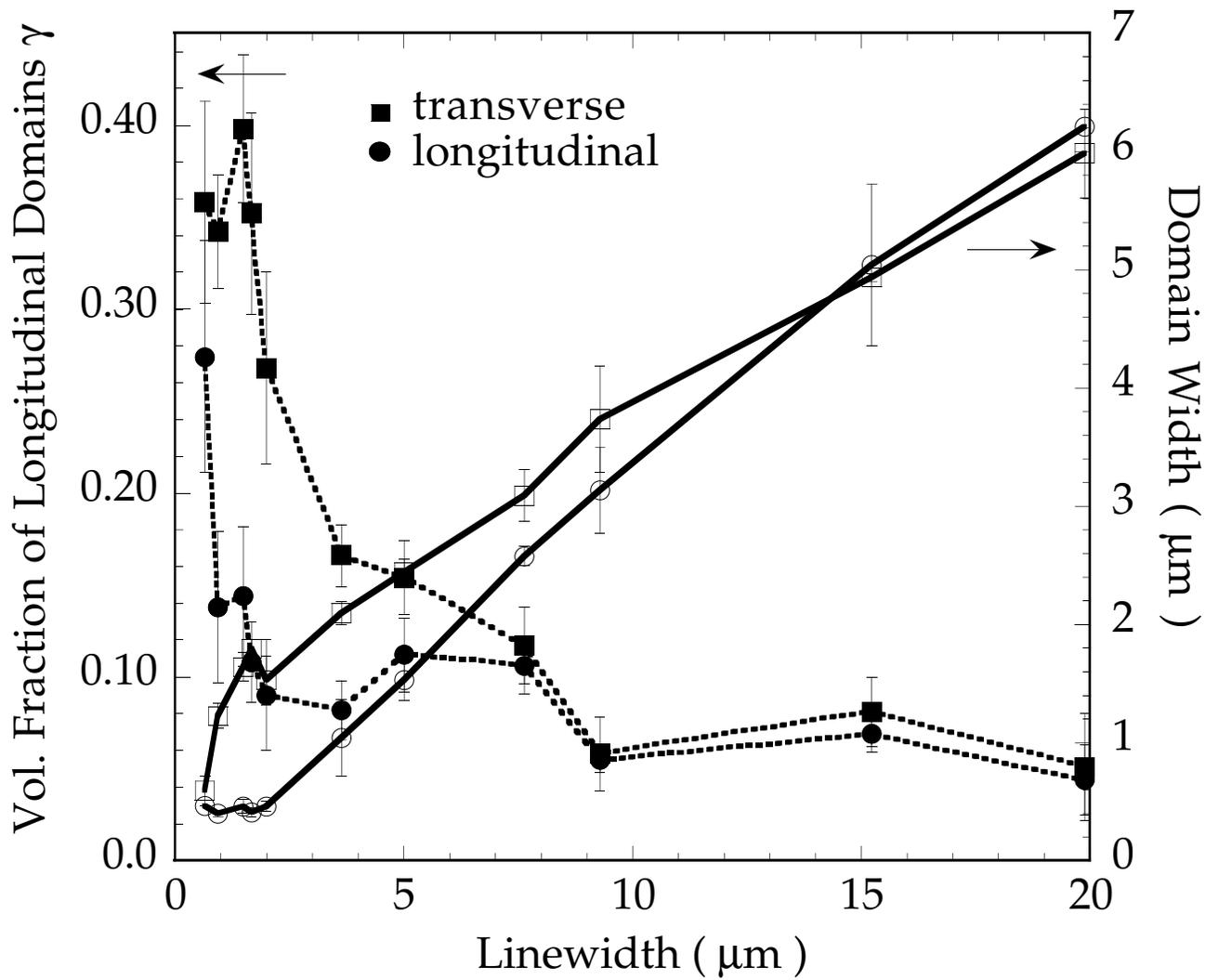
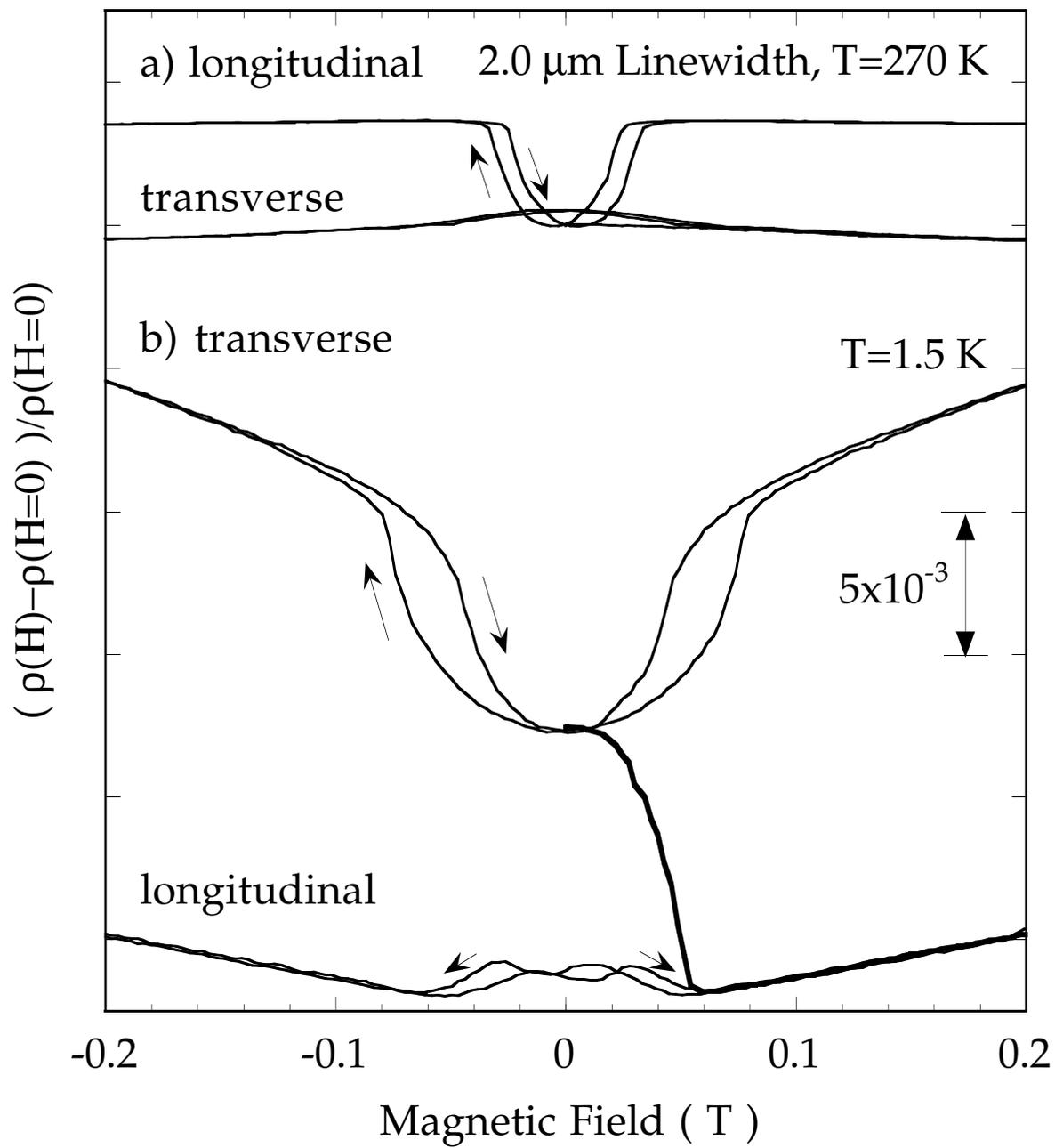
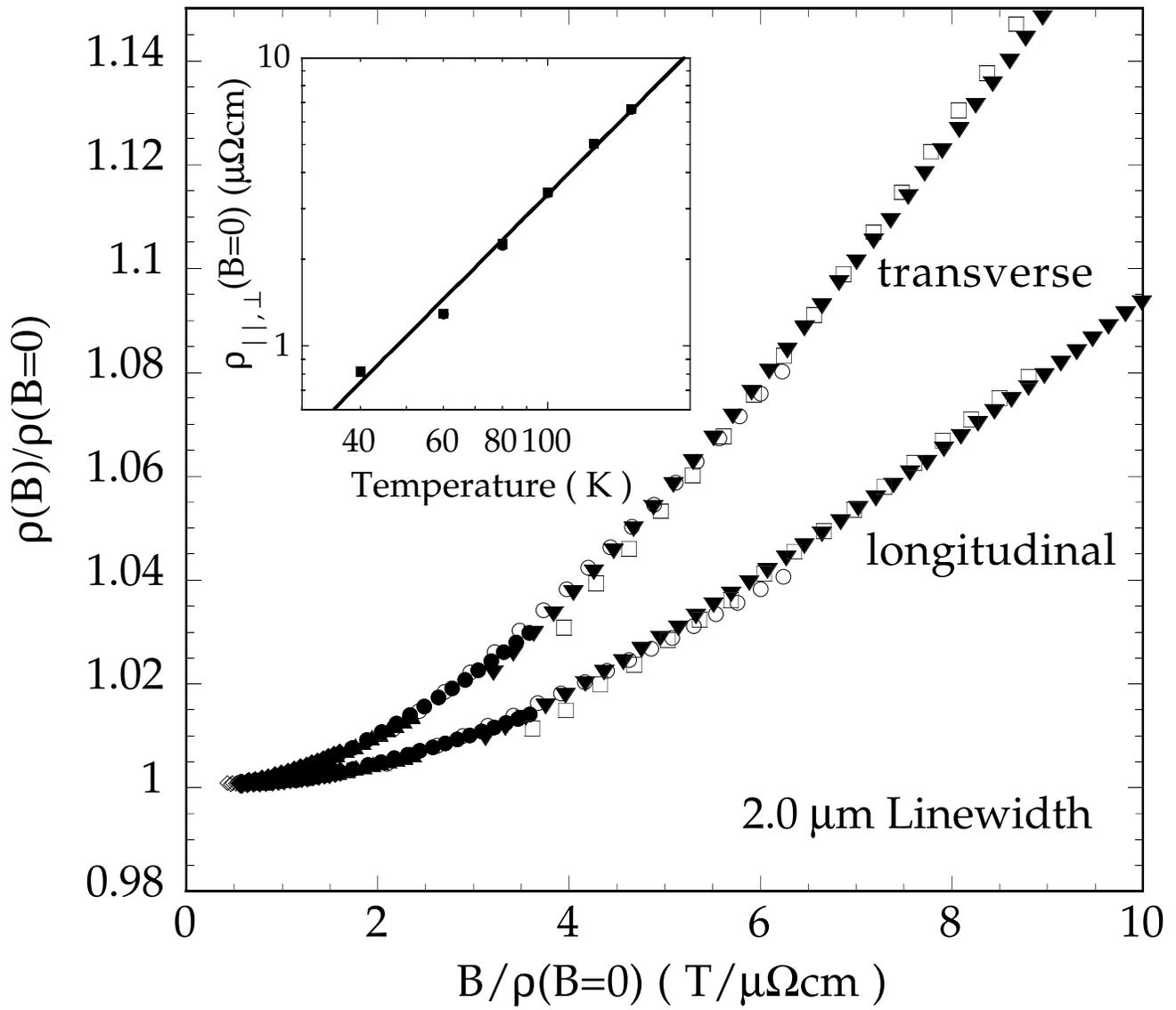
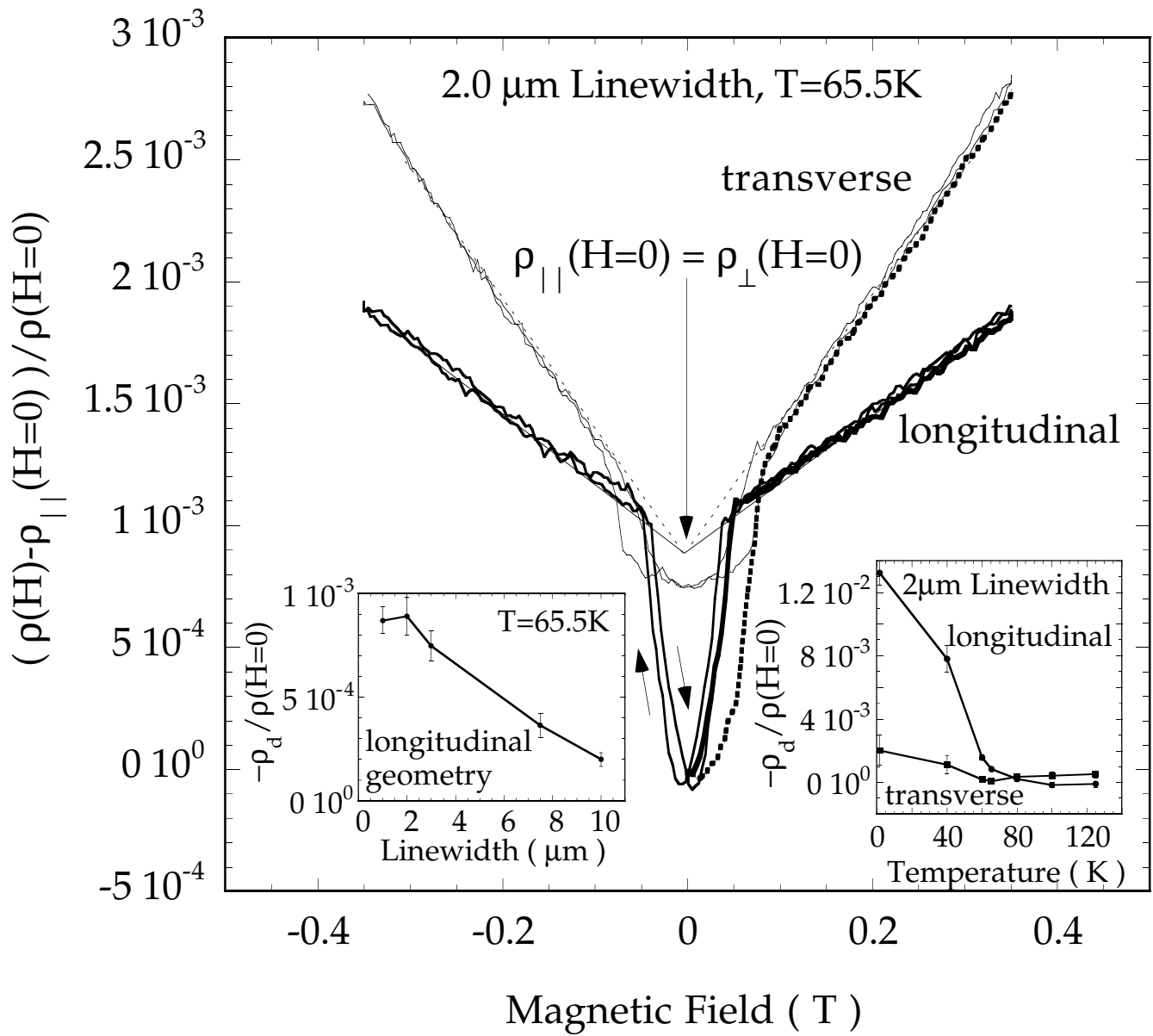
\end{document}